\documentclass[aps,prl,showpacs,twocolumn,groupedaddress]{revtex4}  
\usepackage{graphicx}  
\usepackage{dcolumn}   
\usepackage{bm}        
\usepackage{amssymb}   
\usepackage{multirow}
\usepackage{epsfig}

\newcommand{\zgamma}{\ensuremath{Z/\gamma^{*}}}
\newcommand{\stwcenter}{0.2321}
\newcommand{\stwcentercorr}{0.2326} 
\newcommand{\stwstaterr}{0.0018}
\newcommand{\stwsysterr}{0.0006}
\newcommand{\stwsysterrpdf}{0.0005}
\newcommand{\stwsysterrscale}{0.0003}

\newcommand{\stw}{\ensuremath{\sin^2 \theta_{W}}}
\newcommand{\stweff}{\ensuremath{\sin^2 \theta^{\text{eff}}_{W}}}
\newcommand{\metpaul}{\mbox{$\rlap{\kern0.15em/}E_T$}}
\begin{document}

\title{
 Measurement of the forward-backward charge asymmetry and extraction of $\stweff$ in $p\bar{p} \rightarrow Z/\gamma^{*}+X \rightarrow e^+e^-+X$ events produced at $\sqrt{s}=1.96$ TeV
}
%
\author{V.M.~Abazov$^{36}$}
\author{B.~Abbott$^{75}$}
\author{M.~Abolins$^{65}$}
\author{B.S.~Acharya$^{29}$}
\author{M.~Adams$^{51}$}
\author{T.~Adams$^{49}$}
\author{E.~Aguilo$^{6}$}
\author{S.H.~Ahn$^{31}$}
\author{M.~Ahsan$^{59}$}
\author{G.D.~Alexeev$^{36}$}
% ...

\affiliation{\vspace{0.1 in}(The D\O\ Collaboration)\vspace{0.1 in}}
% ...
\affiliation{$^{6}$University of Alberta, Edmonton, Alberta, Canada,
                Simon Fraser University, Burnaby, British Columbia, Canada,
                York University, Toronto, Ontario, Canada, and
                McGill University, Montreal, Quebec, Canada}
% ...
\affiliation{$^{29}$Tata Institute of Fundamental Research, Mumbai, India}
% ...
\affiliation{$^{31}$Korea Detector Laboratory, Korea University, Seoul, Korea}
% ...
\affiliation{$^{36}$Joint Institute for Nuclear Research, Dubna, Russia}
% ...
\affiliation{$^{49}$Florida State University, Tallahassee, Florida 32306, USA}
% ...
\affiliation{$^{51}$University of Illinois at Chicago,
                Chicago, Illinois 60607, USA}
% ...
\affiliation{$^{59}$Kansas State University, Manhattan, Kansas 66506, USA}
% ...
\affiliation{$^{65}$Michigan State University,
                East Lansing, Michigan 48824, USA}
% ...
\affiliation{$^{75}$University of Oklahoma, Norman, Oklahoma 73019, USA}
% ...
  
\date{April 20, 2008}

\begin{abstract}
We present a measurement of the forward-backward charge asymmetry ($A_{FB}$)
in $p\bar{p} \rightarrow Z/\gamma^{*}+X \rightarrow e^+e^-+X$ events at
a center-of-mass energy of 1.96 TeV using 1.1 fb$^{-1}$ of data collected
with the D0 detector at the Fermilab Tevatron collider.
$A_{FB}$ is measured as a function of the invariant mass of the electron-positron
pair, and found to be consistent with the standard model prediction.
We use the $A_{FB}$ measurement to extract the effective weak mixing angle
$\stweff = \stwcentercorr \pm \stwstaterr ~(\mbox{stat.}) \pm \stwsysterr ~(\mbox{syst.})$.
\end{abstract}
\maketitle

In the standard model (SM), the neutral-current couplings of the $Z$
bosons to fermions ($f$) at tree level are defined as
\begin{equation}
-i \frac{g}{2\cos\theta_W} \cdot \bar{f} \gamma^{\mu}(g_{V}^f - g_{A}^f \gamma_5) f \cdot Z_{\mu}
\end{equation}
where $\theta_{W}$ is the weak mixing angle, and $g_V^{f}$ and $g_A^{f}$ are the
vector and axial-vector couplings with $g_V^f = I_3^f - 2Q_f\stw$ and $g_A^f = I_3^f$.
Here $I_3^f$ is the weak isospin component of the fermion and $Q_f$ its charge.
The presence of both vector and axial-vector couplings
in $q\bar{q} \rightarrow \zgamma \rightarrow \ell^+\ell^-$
gives rise to an asymmetry in the polar angle ($\theta$) of the negatively charged
lepton momentum relative to the incoming quark momentum
in the rest frame of the lepton pair. The angular differential cross
section can be written as
\begin{equation}
\frac{d\sigma}{d \cos \theta} = A(1+\cos^{2} \theta) + B\cos\theta,
\end{equation}
where $A$ and $B$ are functions dependent on $I_3^{f}$, $Q_f$, and $\stw$.
Events with $\cos \theta>0$ are called forward events,
and those with $\cos \theta<0$ are called backward events.

The forward-backward charge asymmetry, $A_{FB}$, is defined as
\begin{equation}
 A_{FB} = \frac{\sigma_{F}-\sigma_{B}}{\sigma_{F}+\sigma_{B}},
\label{form:AFB}
\end{equation}
where $\sigma_{F/B}$ is the integral cross section in the forward/backward configuration.
We measure $A_{FB}$ as a function of the invariant mass of the lepton pair.
To minimize the effect of the unknown transverse momenta of the incoming
quarks in the measurement of the forward and backward cross sections, we
use $\theta$ calculated in the Collins-Soper reference frame \cite{cs_frame}.
In this frame, the polar axis is defined as the
bisector of the proton beam momentum and the negative of the anti-proton beam
momentum when they are boosted into the rest frame of the lepton pair.

The forward-backward asymmetry is sensitive to $\stweff$, which is an effective
parameter that includes higher order corrections. The current world average
value of $\stweff$ at the $Z$-pole is $0.23149 \pm 0.00013$ \cite{pdg}.
Two $\stweff$ measurements are more than two standard deviations
from the world average value: that from the charge asymmetry for $b$ quark
production ($A_{FB}^{0,b}$) from the LEP and SLD collaborations~\cite{lep_sinthetaW}
and that from neutrino and antineutrino cross sections from the NuTeV collaboration~\cite{nutev_sinthetaW}.
The $A_{FB}^{0, b}$ measurement is sensitive to the couplings of
$b$ quarks to the $Z$ boson, and the NuTeV measurement is sensitive
to the couplings of $u$ and $d$ quarks to the $Z$ boson, as is the
measurement presented here. Previous direct measurements of $u$ and $d$ quark
couplings to the $Z$ are of limited precision~\cite{cdf_RunII, H1}. Thus, modifications to the SM
that would affect only $u$ and $d$ couplings are poorly constrained.
In addition, $A_{FB}$ measurements at the Tevatron can be
performed up to values of the dilepton mass exceeding those
achieved at LEP and SLC, therefore becoming sensitive to
possible new physics effects~\cite{zprime, led}. Although direct
searches for these new phenomena in the $\zgamma \rightarrow \ell^+\ell^-$ final
state have been recently performed by the CDF and D0
collaborations~\cite{highmass}, charge asymmetry measurements are sensitive to
different combination of couplings, and can provide complementary
information~\cite{highmass_CDF}.

The CDF collaboration measured $A_{FB}$ using 108 pb$^{-1}$ of data
in Run I~\cite{cdf_RunI} and 72 pb$^{-1}$ of data in Run II~\cite{cdf_RunII}.
This analysis uses $1066 \pm 65$ pb$^{-1}$ of data~\cite{d0lumi}
collected with the D0 detector~\cite{d0det} at the Fermilab Tevatron
collider at a center-of-mass energy of 1.96 TeV to measure the $A_{FB}$
distribution and extract $\stweff$.

\indent To select $\zgamma$ events, we require two
isolated electromagnetic (EM) clusters that have shower shapes
consistent with that of an electron. EM candidates are required to
have transverse momentum $p_T>25 ~\mbox{GeV}$.
The dielectron pair must have a reconstructed invariant mass
$50<M_{ee}<500 ~\mbox{GeV}$. If an event has both its EM candidates
in the central calorimeter (CC events), each must be
spatially matched to a reconstructed track in the tracking system.
Because the tracking efficiency decreases with magnitude of the rapidity in
the end calorimeter, events with one candidate in the central and one candidate in the end
calorimeter (CE events) are required to have a matching track only for that in the
central calorimeter. For CC events, the two candidates are further required to have
opposite charges. For CE events, the determination of forward or
backward is made according to the charge of the EM candidate in the central calorimeter.
A total of 35,626 events remain after application of all selection criteria,
with 16,736 CC events and 18,890 CE events.
The selection efficiencies are measured using $Z/\gamma^{*} \rightarrow ee$ data with
the tag-probe method \cite{tag-probe}, and
no differences between forward and backward events are observed.

The asymmetry is measured in 14 $M_{ee}$ bins within the $50<M_{ee}<500$ GeV range.
The bin widths are determined by the mass resolution, of order $(3 - 4)\%$, and event statistics. 

Monte Carlo (MC) samples for the $\zgamma \rightarrow e^{+}e^{-}$ process
are generated using the {\sc pythia} event generator~\cite{pythia} using
the CTEQ6L1 parton distribution functions (PDFs)~\cite{cteq}, followed by
a detailed {\sc geant}-based simulation of the D0 detector~\cite{geant}.
To improve the agreement between data and simulation,
selection efficiencies determined by the MC are corrected to corresponding
values measured in the data.
Furthermore, the simulation is tuned to reproduce the calorimeter
energy scale and resolution, as well as the distributions of the
instanteneous luminosity and $z$ position of the event primary
vertex observed in data. Next-to-leading order (NLO) quantum chromodynamics
(QCD) corrections for $\zgamma$ boson production~\cite{resbos, NLO_corr} are
applied by reweighting the $\zgamma$ boson transverse momentum, rapidity,
and invariant mass distributions from {\sc pythia}.

The largest background arises from photon+jets and multijet final states in which
photons or jets are mis-reconstructed as electrons.
Smaller background contributions arise from electroweak processes
that produce two real electrons in the final state.
The multijet background is estimated using collider data by fitting
the electron isolation distribution in data to the sum of the
isolation distributions from a pure electron sample and
an EM-like jet sample. The pure electron sample is obtained
by enforcing tighter track matching requirements on the two electrons with
$80<M_{ee}<100$ GeV. The EM-like jets sample is obtained from a sample where only
one good EM cluster and one jet are back-to-back in azimuthal angle $\phi$.
The contamination in the EM-like jets sample from $W \rightarrow e\nu$
events is removed by requiring missing transverse energy $\metpaul<10$ GeV.
The average multijet background fraction over the entire mass region is found to be approximately $0.9\%$.
Other SM backgrounds due to $W+\gamma$,
$W+$jets, $WW$, $WZ$ and $t\bar{t}$ are estimated
separately for forward and backward events using {\sc pythia} events passed through the {\sc geant} 
simulation. Higher order corrections to the {\sc pythia} leading order (LO)
cross sections have been applied~\cite{NLO_corr, WW_NLO_corr, ttbar_NLO_corr}.
These SM backgrounds are found to be negligible for almost all mass bins.
The $\zgamma \rightarrow \tau^+\tau^-$ contribution is similarly negligible.

In the SM, the $A_{FB}$ distribution is fully determined by the
value of $\stweff$ in a LO prediction for the process
$q\bar{q} \rightarrow \zgamma \rightarrow \ell^+\ell^-$. The value of $\stweff$ is
extracted from the data by comparing the background-subtracted
raw $A_{FB}$ distribution with templates corresponding to different
input values of $\stweff$ generated with {\sc pythia} and {\sc geant}-based MC simulation.
Although $\stweff$ varies over the full mass range $50<M_{ee}<500$ GeV, it is
nearly constant over the range $70<M_{ee}<130$ GeV. Over this region, we measure
$\stweff =\stwcenter \pm \stwstaterr ~(\mbox{stat.}) \pm \stwsysterr~(\mbox{syst.})$.
The primary systematic uncertainties are due to the PDFs
(\stwsysterrpdf) and the EM energy scale and resolution (\stwsysterrscale).
We include higher order QCD and electroweak corrections using
the {\sc zgrad2}~\cite{zgrad} program with the
generator-level $\zgamma$ boson $p_T$ distribution tuned to match
our measured distribution~\cite{zpt}. The effect of higher order corrections
results in a central value of $\stweff=0.2326$~\cite{explaination}.

Due to the detector resolution, events may be reconstructed in a different mass bin
than the one in which they were generated.
The CC and CE raw $A_{FB}$ distributions are unfolded separately and then combined.
The unfolding procedure is based on an iterative application of the method of
matrix inversion~\cite{matrix_inversion}.
A response matrix is computed as $R_{ij}^{FF}$ for an event that is measured as forward in $M_{ee}$
bin $i$ to be found as forward and in bin $j$ at the generator level.
Likewise, we also calculate the response matrices for backward
events being found as backward ($R_{ij}^{BB}$), forward as backward
($R_{ij}^{FB}$), and backward as forward ($R_{ij}^{BF}$).
Four matrices are calculated from the {\sc geant} MC simulation and used to unfold the raw
$A_{FB}$ distribution. The method was verified by comparing the true and unfolded spectrum
generated using pseudo-experiments.

The data are further corrected for acceptance and selection
efficiency using the {\sc geant} simulation. The overall
acceptance times efficiency rises from $3.5\%$ for $50<M_{ee}<60$ GeV
to $21\%$ for $250<M_{ee}<500$ GeV.

The electron charge measurement in the central calorimeter determines
whether an event is forward or backward.
Any mismeasurement of the charge of the electron results in a dilution of $A_{FB}$.
The charge misidentification rate, $f_Q$, is measured using {\sc geant}-simulated
$\zgamma \rightarrow e^+e^-$ events tuned to the average rate measured in data. The
misidentification rate rises from 0.21\% at $50<M_{ee}<60$ GeV to 0.92\% at $250<M_{ee}<500$ GeV.
The charge misidentification rate is included as a dilution factor
$\cal{D}$ in $A_{FB}$, with ${\cal{D}}=(1-2f_Q)/(1-2f_Q+f^2_Q)$ for CC events
and ${\cal{D}}=(1-2f_Q)$ for CE events.

\begin{center}
\begin{figure}[htbp]
\epsfig{file=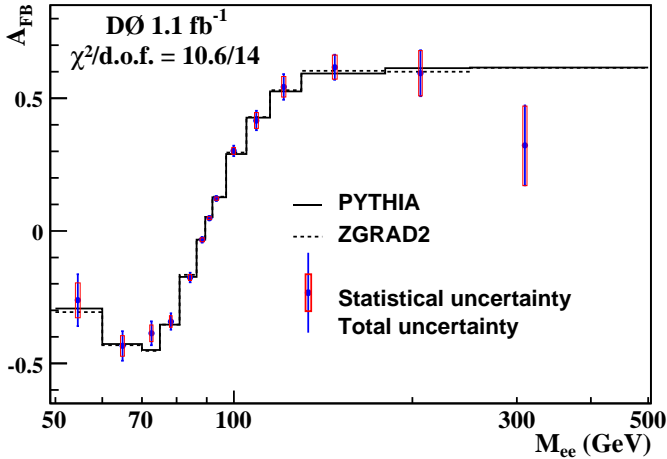, scale=0.45}
\caption{\small Comparison between the unfolded $A_{FB}$ (points) and
the {\sc pythia} (solid curve) and {\sc zgrad2} (dashed line)
predictions. The inner (outer) vertical lines
show the statistical (total) uncertainty.}
\label{fig:compare_afb}
\end{figure}
\end{center}

\indent The final unfolded $A_{FB}$ distribution using both CC and CE events is shown in 
Fig.~\ref{fig:compare_afb}, compared to the {\sc pythia} prediction using the CTEQ6L1 PDFs~\cite{cteq} 
and the {\sc zgrad2} prediction using the CTEQ5L PDFs~\cite{cteq5}.
The $\chi^2/\mbox{d.o.f.}$ with respect to the {\sc pythia} prediction is $16.1/14$ for CC,
$8.5/14$ for CE, and $10.6/14$ for CC and CE combined.
The systematic uncertainties for the unfolded $A_{FB}$ distribution
arise from the electron energy scale and resolution,
backgrounds, limited MC samples used to calculate the response matrices,
acceptance and efficiency corrections, charge misidentification and PDFs.
The unfolded $A_{FB}$ together with the {\sc pythia} and {\sc zgrad2} predictions
for each mass bin can be found in Table~\ref{Tab:afb_final}.
The correlations between invariant mass bins are shown in Table~\ref{Tab:corr_matrix}.

\indent In conclusion, we have measured the forward-backward charge
asymmetry for the $p\bar{p} \rightarrow \zgamma+X \rightarrow e^{+}e^{-}+X$
process in the dielectron invariant mass range 50 -- 500 GeV using
1.1~fb$^{-1}$ of data collected by the D0 experiment.
The measured $A_{FB}$ values are in good agreement with the SM predictions.
We use the $A_{FB}$ measurements in the range $70<M_{ee}<130$ GeV to determine
$\stweff = \stwcentercorr \pm \stwstaterr~(\mbox{stat.}) \pm \stwsysterr~(\mbox{syst.})$.
The precision of this measurement is comparable to that obtained from LEP measurements
of the inclusive hadronic charge asymmetry~\cite{lep_sinthetaW} and that of NuTeV
measurement~\cite{nutev_sinthetaW}. Our measurements of $\stweff$ in a dilepton mass
region dominated by $Z$ exchange, which is primarily sensitive to the vector
coupling of the $Z$ to the electron, and of $A_{FB}$ over a wider mass region,
which is in addition sensitive to the couplings of the $Z$ to light quarks,
agrees well with predictions. With about 8~fb$^{-1}$ of data expected by the end of Run II,
a combined measurement of $A_{FB}$ by the CDF and D0 collaborations using electron and muon
final states could lead to a measurement of $\stweff$ with
a precision comparable to that of the current world average.
Further improvements to current MC generators, incorporating higher order QCD
and electroweak corrections, would enable the use of such measurement in a global
electroweak fit.

\begin{table}[!htb]
\begin{center}
\begin{tabular}{r@{$\,- \,$}lcccc} \hline \hline
\multicolumn{2}{c}{$M_{ee}$ range}  & $\langle M_{ee} \rangle$  & \multicolumn{2}{c}{\ Predicted $A_{FB}$}  &  \ \ \multirow{2}{*}{Unfolded $A_
{FB}$} \\
\multicolumn{2}{c}{(GeV)} &  {(GeV)}  & {\ \sc pythia} & {\sc zgrad2} &  \\  \hline
50 & 60 & 54.5   & $-0.293$ & $-0.307$  &  \ \ $-0.262 \pm 0.066 \pm  0.072$ \\
60 & 70 & 64.9   & $-0.426$ & $-0.431$  &  \ \ $-0.434 \pm 0.039 \pm  0.040$ \\
70 & 75 & 72.6   & $-0.449$ & $-0.452$  &  \ \ $-0.386 \pm 0.032 \pm  0.031$ \\
75 & 81 & 78.3   & $-0.354$ & $-0.354$  &  \ \ $-0.342 \pm 0.022 \pm  0.022$ \\
81 & 86.5 & 84.4   & $-0.174$ & $-0.166$  &  \ \ $-0.176 \pm 0.012 \pm  0.014$ \\
86.5 & 89.5 & 88.4   & $-0.033$ & $-0.031$  &  \ \ $-0.034 \pm 0.007 \pm  0.008$ \\
89.5 & 92 & 90.9   & \ \ $0.051$ & \ \ $0.052$  & \ \ \ \ $0.048 \pm 0.006 \pm  0.005$ \\
92 & 97 & 93.4   & \ \ $0.127$ & \ \ $0.129$  & \ \ \ \ $0.122 \pm 0.006 \pm  0.007$ \\
97 & 105 & 99.9   & \ \ $0.289$ & \ \ $0.296$  & \ \ \ \ $0.301 \pm 0.013 \pm  0.015$ \\
105 & 115 & 109.1  & \ \ $0.427$ & \ \ $0.429$  & \ \ \ \ $0.416 \pm 0.030 \pm  0.022$ \\
115 & 130 & 121.3  & \ \ $0.526$ & \ \ $0.530$  & \ \ \ \ $0.543 \pm 0.039 \pm  0.028$ \\
130 & 180 & 147.9  & \ \ $0.593$ & \ \ $0.603$  & \ \ \ \ $0.617 \pm 0.046 \pm  0.013$ \\
180 & 250 & 206.4  & \ \ $0.613$ & \ \ $0.600$  & \ \ \ \ $0.594 \pm 0.085 \pm  0.016$ \\
250 & 500 & 310.5  & \ \ $0.616$ & \ \ $0.615$  & \ \ \ \ $0.320 \pm 0.150 \pm  0.018$ \\ \hline
\hline
\end{tabular}
\caption{The first column shows the mass ranges used. The second column shows the
cross section weighted average of the invariant mass in each mass bin derived from {\sc pythia}. The
third and fourth columns show the $A_{FB}$ predictions from {\sc pythia} and {\sc zgrad2}.
The last column is the unfolded $A_{FB}$; the
first uncertainty is statistical, and the second is systematic.}
\label{Tab:afb_final}
\end{center}
\end{table}

\begin{table*}
\begin{ruledtabular}
\begin{tabular}{c|cccccccccccccc}
Mass bin & 1 & 2 & 3 & 4 & 5 & 6 & 7 & 8 & 9 &10 & 11 & 12 & 13 & 14 \\ \hline
        1 &      1.00 &      0.21 &      0.04 &     0.00 &     0.00 &      0.01 &      0.00 &      0.01 &      0.00 &      0.00 &      0.00 &  0.00 &     0.00 &     0.00 \\
        2 &      &      1.00 &      0.42 &      0.08 &      0.02 &      0.01 &      0.02 &      0.01 &      0.00 &      0.00 &      0.00 & 0.00 &     0.00 &     0.00 \\
        3 &      &      &      1.00 &      0.49 &      0.13 &      0.04 &      0.03 &      0.02 &      0.01 &      0.00 &      0.00 &      0.00 &     0.00 &      0.00 \\
        4 &      &      &      &      1.00 &      0.52 &      0.16 &      0.08 &      0.04 &      0.01 &      0.00 &      0.00 &     0.00 & 0.00 &      0.00 \\
        5 &      &      &      &      &      1.00 &      0.72 &      0.32 &      0.11 &      0.01 &      0.00 &      0.00 &      0.00 &      0.00 &      0.00 \\
        6 &      &      &      &      &      &      1.00 &      0.79 &      0.40 &      0.03 &      0.00 &     0.00 &      0.00 &      0.00 &   0.00 \\
        7 &      &      &      &      &      &      &      1.00 &      0.80 &      0.15 &      0.01 &      0.00 &      0.00 &     0.00 &     0.00 \\
        8 &      &      &      &      &      &      &      &      1.00 &      0.50 &      0.04 &      0.00 &     0.00 &      0.01 &      0.00 \\
        9 &      &      &      &      &      &      &      &      &      1.00 &      0.38 &      0.04 &      0.00 &      0.00 &      0.00 \\
       10 &      &      &      &      &      &      &      &      &      &      1.00 &      0.30 &      0.01 &      0.00 &      0.00 \\
       11 &      &      &      &      &      &      &      &      &      &      &      1.00 &      0.14 &      0.00 &      0.00 \\
       12 &      &      &      &      &      &      &      &      &      &      &      &      1.00 &      0.06 &      0.00 \\
       13 &      &      &      &      &      &      &      &      &      &      &      &      &      1.00 &      0.06 \\
       14 &      &      &      &      &      &      &      &      &      &      &      &      &      &      1.00 \\
\end{tabular}
\caption{Correlation coefficients between different $M_{ee}$ mass bins. Only half of the symmetric correlation matrix is presented.}
\label{Tab:corr_matrix}
\end{ruledtabular}
\end{table*}

%
We thank the staffs at Fermilab and collaborating institutions,
and acknowledge support from the
DOE and NSF (USA);
CEA and CNRS/IN2P3 (France);
FASI, Rosatom and RFBR (Russia);
CNPq, FAPERJ, FAPESP and FUNDUNESP (Brazil);
DAE and DST (India);
Colciencias (Colombia);
CONACyT (Mexico);
KRF and KOSEF (Korea);
CONICET and UBACyT (Argentina);
FOM (The Netherlands);
STFC (United Kingdom);
MSMT and GACR (Czech Republic);
CRC Program, CFI, NSERC and WestGrid Project (Canada);
BMBF and DFG (Germany);
SFI (Ireland);
The Swedish Research Council (Sweden);
CAS and CNSF (China);
and the
Alexander von Humboldt Foundation.


\begin{thebibliography}{99}
\bibitem{explaination}
  This value of $\stweff$ cannot be compared directly with the world average due to the different treatment of electroweak corrections.
\end{thebibliography}
\end{document}